\documentclass[twocolumn,showpacs,showkeys,footinbib,prb,preprintnumbers,amsmath,amssymb]{revtex4}

\usepackage{graphicx}
\usepackage{graphics}
\usepackage{dcolumn}
\usepackage{bm}

\begin{document}

\title{Dynamics of twist glass transition of a $\pi$-conjugated polymer investigated by NMR relaxation spectroscopy: poly({\it para}-phenylene)}

\author{Naoki Asakawa}
\thanks{Corresponding Author}
\email{nasakawa@bio.titech.ac.jp}
\affiliation{Department of Biomolecular Engineering, 
Tokyo Institute of Technology.  
4259 B-55 Nagatsuta-cho, Midori-ku, Yokohama, Kanagawa 226-8501, Japan}%

\author{Manabu Ohira}
\affiliation{Department of Biomolecular Engineering, 
Tokyo Institute of Technology.  
4259 B-55 Nagatsuta-cho, Midori-ku, Yokohama, Kanagawa 226-8501, Japan}

\author{Koji Yazawa}
\affiliation{Department of Biomolecular Engineering, 
Tokyo Institute of Technology.  
4259 B-55 Nagatsuta-cho, Midori-ku, Yokohama, Kanagawa 226-8501, Japan}

\author{Takakazu Yamamoto}%
\affiliation{Chemical Resources Laboratory, 
Tokyo Institute of Technology.  
4259 Nagatsuta-cho, Midori-ku, Yokohama, Kanagawa 226-8503, Japan}

\author{Yoshio Inoue}
\affiliation{Department of Biomolecular Engineering, 
Tokyo Institute of Technology. 
4259 B-55 Nagatsuta-cho, Midori-ku, Yokohama, Kanagawa 226-8501, Japan}

\date{\today}

\begin{abstract}
Dynamics of Yamamoto-type poly({\it para}-phenylene)[PPP] was investigated by differential scanning calorimetry(DSC) and proton solid-state NMR relaxation spectroscopy.
The DSC chart shows the baseline jump without latent heat at 295K,
which is due to the glass transition of the polymer. 
From the variable temperature proton longitudinal relaxation time($T_1$)
measurements,
relatively short $T_1$ is observed over the wide temperatures range from 250K
(closed to Vogel-Fulcher-Tamman temperature) to 360K, 
inferred the existence of cooperative critical slowing down associated with 
the glass transition.
The frequency dependence of proton longitudinal relaxation time at 295K
shows $R_1 \sim \omega^{-0.5}$ dependence,
which is due to the one-dimensional diffusion-like motion of the backbone
conformational modulation.
The frequency dependence is held at least up to 360K.
From these experiments, we were able to observe the twist glass transition of 
the backbone of PPP,
of which critical dynamics has a universality class of the three-dimensional XY model.
\end{abstract}



\maketitle


\section{Introduction}
"Conformon" is known as one of elementary excitations concerning molecular structure.\cite{Andrew}
The quasiparticle in $\pi$-conjugated polymers has been of particular interest
because static and dynamical aspects of the quasiparticle can closely be related
to the electronic and optical properties of the polymers.
Poly({\it para}-phenylene)[PPP] is one of the polymers on which extensive studies have been
conducted for physical properties,
because of its potential functionalities such as
electrical conduction or optical applications
(electrical conductivity \cite{Shacklette1979},
rechargeable battery \cite{Shacklette1982,Satoh1986},
electrorheology \cite{Shiga,Plocharski1997,Plocharski1999,Sim,Sohn2002_JMS,Sohn2002_JAPS},
electroluminescence \cite{Grem1992_AM,Grem1992_SM}).
In spite of extensive studies,
however,
even molecular and crystal structures have not been established yet.

Particular attention has been devoted to the phase transition of
$\pi$-conjugated small molecules which have the incommensurate (IC)
phase
(such as biphenyl \cite{Liu,vonLaue},
dichlorobiphenylsulphone \cite{Souza},
etc.).
In these phases,
disordered crystals show the long range breaking of the lattice
periodicity.
Therefore the wavelength of the modulation is not an integral multiple
of the unit cell,
which means that translational symmetry is lost.
In most incommensurate crystals,
power-law temperature and frequency dependences on $^1$H longitudinal
relaxation time were observed.\cite{Liu,Souza}
Furthermore,
the relaxation rate rises dramatically around the transition temperature
$T_i$.
This is due to the fact that the longitudinal relaxation is controlled
by low frequency excitations of the collective phase modes (phasons)
of the order parameter(i.e., backbone twist) corresponding to the critical slowing down of molecular motion
accompanying the phase transition.\cite{Liu,Souza,Decker}

As far as $\pi$-conjugated polymers are concerned,
although it is well known that $\pi$-conjugated polymers shows
discommensuration(so-called soliton) by carrier doping,\cite{Winokur}  
there have never been reported to confirm the existence of
incommensurate modulations as a purely, molecular structural problem.
Recently, we have investigated the dynamics of the conformon in
a regio-regulated $\pi$-conjugated polymer and found the existence of 
modulation waves of backbone twist in its quasiordered phase which coexists with the crystalline phase in association with the order-disorder phase transition.\cite{Asakawa}
Although it is still unclear whether the modulation waves are incommensurate,
the waves show anomalous dispersion of which the group velocity is approximately 
10$^7$ time larger than the phase velocities.

In this article,
we shall give attention to the conformon in a $\pi$-conjugated polymer glass
and investigate the critical dynamics of twist glass transition in poly({\it para}-phenylene)[PPP] mainly by nuclear magnetic resonance(NMR) relaxation spectroscopy, which deal with the frequency dependence of NMR relaxation rate in order to obtain information about dimensionality of the correlation function or spectral density function of fluctuations of local fields in a sample.
So far various glassy polymers have received much attention because of their unique 
visco-elastic properties. 
While the great amount of publications concerning glassy polymers have been devoted to
conventional engineering plastics or elastomers with a flexible backbone,\cite{Richter1988,Kanaya,Sokolov1993,Rossler,Dob1998}
few publications have been available for dynamics of glass transition 
in needle-like or plate-like polymers with rigid backbone 
such as $\pi$-conjugated polymers.

Among various methods for the preparation of PPP
(the Kovacic method \cite{Kovacic1963},
the Yamamoto method \cite{Yamamoto1977,Yamamoto1978,Yamamoto1990,Yamamoto1992},
the biotechnology-based method \cite{Ballard1983,Ballard1988},
the electrolytic polymerization method \cite{Satoh1985}),
organometallic dehalogenation polycondensation of 1,4-dihalobenzene,
particularly using zero-valent Ni complex \cite{Yamamoto1992},
affords PPP with well-defined linkage between the monomer units({\it para}-form;no branches and cross-linked structures),
and therefore the Yamamoto-type is suitable for detailed investigation 
of dynamics of PPP.
The x-ray diffraction study of the powder sample indicates that 
two molecular chains are packed in an orthorhombic unit cell
(a $=$ 0.779 nm; b $=$ 0.551 nm).\cite{Sasaki}
The molecular conformation is not strictly planar;
namely, there is a twist between adjacent phenylene-ring planes in a molecular alternate
with an angle of approximately 20$^\circ$.


\section{Experiments}


\subsection{Materials}

PPP was prepared by dehalogenation polycondensation of
1,4-dihalobenzene with a zero-valent nickel complex \cite{Yamamoto1992}.
From the conventional element analysis, 
the present PPP contains 2.7 \% bromine.
From the value of bromine content, we estimated an average molecular weight of 2960
(DP (degree of polymerization) $=$ 38) and 5930 (DP $=$ 76),
if one polymer chain has bromine at one terminal unit and at both
terminal units, respectively.


\subsection{DSC Measurements}

DSC thermograms of PPP were recorded on a Seiko DSC 220 system
connected to a SSC5300 workstation.
The heating rate is 10 K/min.
After the first heating scan,
the sample was rapidly quenched by liquid nitrogen and then
heated again at the same rate.


\subsection{Solid-state NMR Measurements}

We carried out variable temperature $^1$H longitudinal
relaxation time ($T_1$) measurements with variable frequency condition.
We used the saturation-recovery method for longitudinal relaxation time
($T_{1H}$) measurements at $^1$H resonance frequencies of 25 MHz
(0.59 Tesla), 270 MHz (6.34 Tesla), and 400 MHz (9.4 Tesla)
and the spin-locking method for longitudinal relaxation
time in the rotating frame of a radio frequency field ($T_{1\rho}$)
measurements at $^1$H resonance frequency of 400 MHz with varying 
temperature.
Varian UnityINOVA400 FT NMR spectrometer (the spectral width of 500 kHz) 
was used for 9.4 Tesla experiments.
JEOL GSX-270 FT NMR spectrometer (the spectral width of 100 kHz) was 
used for 0.59 Tesla and 6.34 Tesla experiments.
For $T_{1\rho}$ measurements on a wide range of radio-frequency (rf)
field,
we made a home-built probe with a solenoid type micro-coil with a
radius of about 1 mm and a length of about 5 mm. 
The intensity of rf field was over the range from 16 kHz to 106 kHz.


\section{Results and Discussion}


\subsection{DSC Measurements}

\begin{figure}[htbp]
\begin{center}
\includegraphics*[scale=0.4]{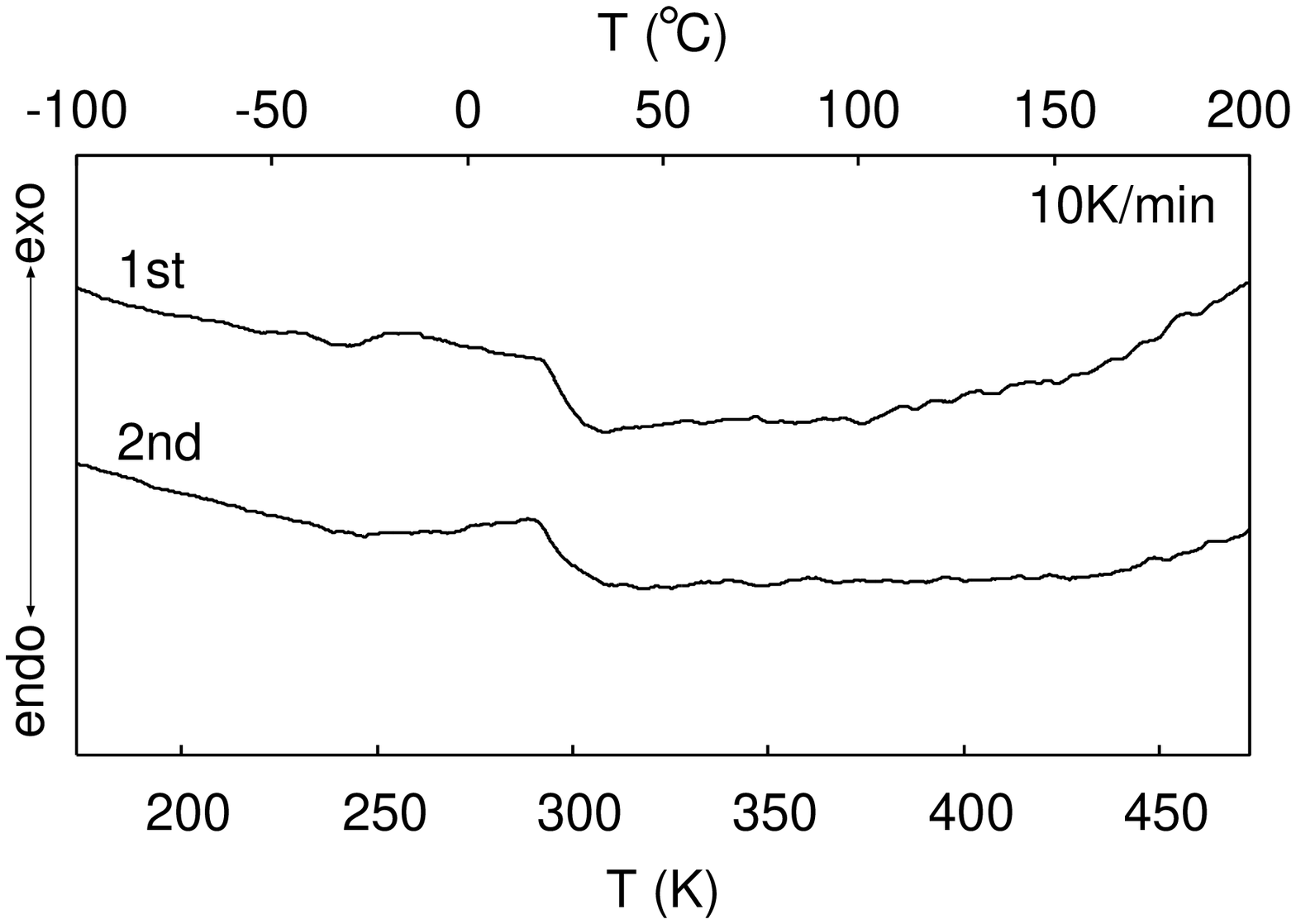}
\caption{DSC charts for a powder sample of PPP. The heating rate was 10 K/min. After the first heating scan, the samples were quenched by liquid nitrogen.}
\label{fig:DSC}
\end{center}
\end{figure}

The thermal property of PPP was examined by DSC (Fig.\ref{fig:DSC}).
The thermograms indicate that for PPP there is a specific heat jump
without latent heat near 295 K.
The jump is due to a phase
transition of the second kind in the crystalline PPP or due to a glass 
transition in the amorphous PPP.

The heat capacities of crystalline PPP oligomers (DP $=$ 3, 4, 5, 6)
have been measured by calorimetry \cite{Saito}.
The thermal anomalies with latent heat were detected and attributed 
to a phase transition,
through which an average molecular conformation changes from the 
planar to the twisted one on cooling.
Such a finite torsional angle results from the balance between the
stabilizing force of conjugation of the phenyl rings and the repulsion
between nearest {\it ortho}-hydrogens,
and this situation induces existence of a double-well potential within
the molecule.
For molecules with larger degrees of polymerization, a twist phase transition of the order-disorder type 
will bring about a thermal anomaly with latent heat.
It can therefore be said that the twist phase transition will not exist in the crystalline portion of PPP.
This is consistent with the fact that the temperature of the thermal anomaly 
we detected(295K) is much lower than that predicted for the polymer by the adiabatic specific heat
measurement(395K$\pm$10K)\cite{Saito}. 

It is widely accepted that PPP is a thermally and thermooxidatively 
stable material \cite{Kovacic1987,Yoshino,Ballard}.
Hence, it seems to be difficult to regard the thermal anomaly as a conventional 
glass transition in the amorphous PPP.
In order to clarify the thermal anomaly, 
we performed the nuclear spin relaxation measurements in the following.


\subsection{$^1$H Spin-lattice Relaxation Time Measurements}
\begin{figure}[htbp]
\begin{center}
\includegraphics*[scale=0.4]{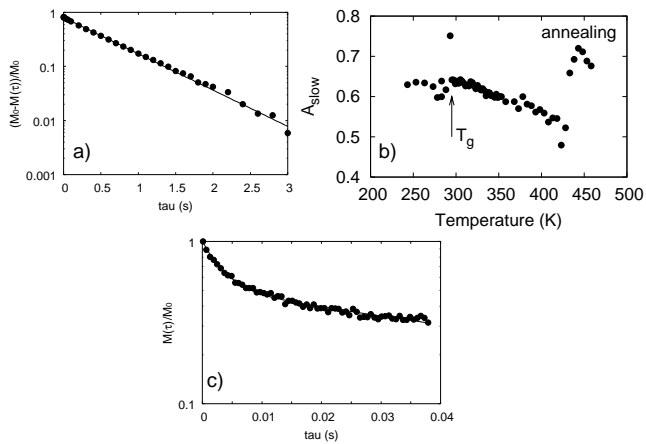}
\caption{A typical saturation recovery profile for PPP obtained by 400 MHz $^1$H longitudinal relaxation time measurements at 295K(a) and the temperature dependence of the fraction of the longer relaxation time(b). The shorter fraction was not able to be detected because of slower analog-digital conversion rate than the relaxation rate.   A typical decay profile for PPP obtained by 400 MHz $^1$H longitudinal relaxation time measurements in the rotating frame at 295K(c)}
\label{fig:T1_fit}
\end{center}
\end{figure}

\begin{figure}[htbp]
\begin{center}
\includegraphics*[scale=0.8]{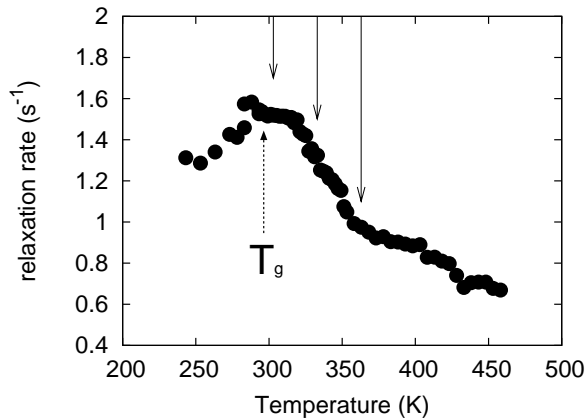}
\caption{The temperature dependence of 400 MHz $^1$H longitudinal relaxation rate. Dotted arrow indicates 295 K. Solid arrows indicate 303K, 333K, and 363K.}
\label{fig:T1H_log}
\end{center}
\end{figure}

Fig.\ref{fig:T1_fit}a shows a typical profile  of the longitudinal magnetization
as a function of recovery time. The conventional saturation recovery method was used.
We observed the signal recovery as a single exponential of the first order, 
but the prefactor was less than unity.
For relaxation with two components, 
the recovery of magnetization is expressed by the following biexponential function. 
\begin{eqnarray}
\frac{M_0 - M(\tau)}{M_0} &=& A_{\rm fast} e^{-\tau / T_1^{\rm f}} + A_{\rm slow} e^{-\tau / T_1^{\rm s}} \label{biexp}
\end{eqnarray}
If one component relaxes too fast to detect by the NMR receiver,
the first term of the r.h.s. of Eq.\ref{biexp} is omitted and 
Eq.\ref{biexp} is reduced to 
\begin{eqnarray}
\frac{M_0 - M(\tau)}{M_0} &=& A_{\rm slow} e^{-\tau / T_1^{\rm s}}.
\end{eqnarray}
Fig.\ref{fig:T1_fit}b shows the variation of the fraction of slower component,
$A_{\rm slow}$, as a function of temperature.
For the temperatures from 295K up to 430K, 
the fraction was decreased as an increase of temperature.
For over 430K,
we detected the sudden jump of $A_{\rm slow}$, probably due to
the annealing effect of the polymer.
In general, 
mobile unpaired electrons in $\pi$-conjugated polymers can affect nuclear spin relaxations while trapped unpaired electrons does not affect the apparent nuclear spin relaxations, because the hyperfine interaction with trapped electron can be too strong to detect the nuclear spin signal by the conventional method.
Further, an increase of temperature causes decrease(increase) of concentration of trapped(mobile or untrapped) unpaired electron in $\pi$-conjugated polymers, and then causes increase of $A_{\rm slow}$ if the relaxation is occurred by electron.
Since the opposite phenomenon was observed in reality,
the slower relaxation is not attributed to the relaxation 
by unpaired electrons,
but due to molecular dynamics of sites without unpaired electron.
By the variable temperature longitudinal relaxation experiments({\it vide infra},Fig.\ref{fig:T1H_log}),
the slower relaxation is found to be due to fluctuations of the local field by molecular dynamics. 
On the other hand, 
the non-observed component with a faster relaxation($T_1^{\rm f} > 10^4$s) may be attributable to 
temporal fluctuation of the hyperfine interaction with the mobile unpaired electron.

Fig.\ref{fig:T1_fit}c shows the typical decay profile for the proton longitudinal relaxation measurements by the conventional spin-locking method.\cite{Ailion}
First,
the decay profile was biexponential, 
which can be originated from two possibilities;
one is spatially heterogeneous distribution of mobile(untrapped) electrons
and the other is structural heterogeneity by the crystalline/amorphous duality.
Since much faster relaxation rate would be expected for $T_{1\rho}$ than $T_1^{\rm f}$($>10^4$s) in a case of relaxation by mobile unpaired electrons,
the first possibility can be ruled out.
For the results of $T_{1\rho}$, 
we assigned the faster as the relaxation derived from the amorphous 
and the slower as from the crystalline portion.
This is consistent with the tendency of the crossover 
from one-dimensional fluctuation in the crystalline portion 
to the higher dimension;namely, 
the relaxation rate becomes almost insensitive to the smaller $\omega$.\cite{Butler}
On the other hand, 
the amorphous portion preserves the one-dimensionality 
in the fluctuation of the local field.
The recovery with the single exponential found in the $T_1$ recovery(Fig.\ref{fig:T1_fit}a) is
due to the $^1$H-$^1$H spin diffusion between the crystalline and amorphous portions;in other words, 
if the spin diffusion were not there the intrinsic relaxation rate for the amorphous portion would be much larger than that for the crystalline portion.

The 400 MHz $^1$H longitudinal relaxation rate ($T_1^{-1}$) as a
function of temperature is shown in Fig.\ref{fig:T1H_log}.
The increase of $T_1^{-1}$ was observed over the wide temperature range from 250K to 360K,
which is due to the critical fluctuation associated with the transition.

\begin{figure*}[htbp]
\begin{center}
\includegraphics*[scale=0.8]{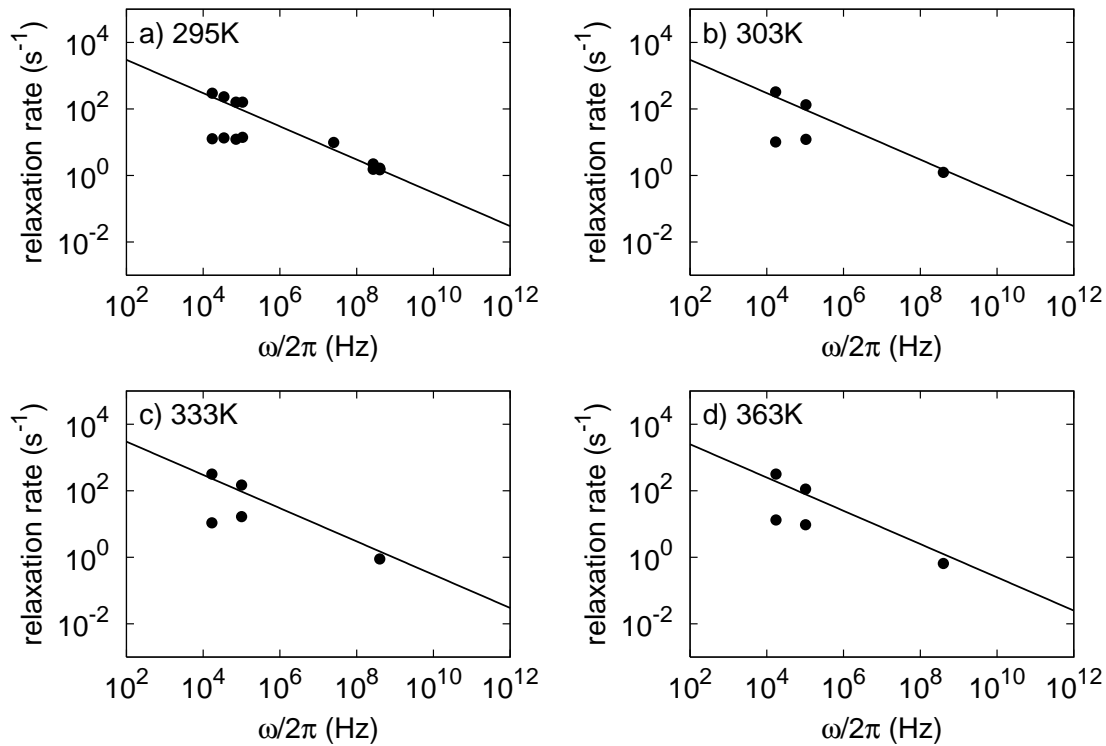}
\caption{The frequency dependence of $^1$H longitudinal relaxation rate at 295K (a), 303K (b), 333K (c), and 363K (d).
The solid lines of the $\omega^{-1/2}$ dependence is shown as a guide for eyes.}
\label{fig:freq_depend_all}
\end{center}
\end{figure*}

The $^1$H longitudinal relaxation rate ($T_1^{-1}$) at 295K
as a function of the frequency is shown in Fig.\ref{fig:freq_depend_all}(a).
The relaxation rate shows the dependence of
$T_1^{-1} \propto \omega^{-0.5}$.
From the fact that the decay profile of the proton longitudinal relaxation
shows the single exponential of the first order(Fig.\ref{fig:T1_fit}), 
we could not straightforwardly attribute the frequency dependence of $T_1^{-1}$ to 
the distribution of correlation time($\tau_c$).
This relationship between $T_1^{-1}$ and $\omega$ is also in disagreement
with that predicted from the classical BPP theory \cite{Bloembergen}.
The BPP theory in which a fluctuation of local field is described as
Lorentzian gives the following spectral density function,
\begin{eqnarray}
J(\omega) &=& \frac{\tau_c}{1+(\omega \tau_c)^2}\\
&\sim& \frac{1}{\omega^2 \tau_c} \quad \text{(if $\omega \tau_c \gg 1$)}.
\end{eqnarray}
In the region where $T_1$ depends on frequency ($\omega \tau_c \gg 1$),
the relation $T_1^{-1} \propto \omega^{-2}$ should be realized.
Therefore an alternative correlation function should be considered in
order to explain the dependence obtained by our experiments.

So far there are several reports that show the $\omega^{-1/2}$
dependence of $T_1^{-1}$.
The ideas of theories are roughly classified into two categories.
First, with assuming existence of an incommensurate phase,
appropriate dynamical susceptibility of the classical damped harmonic
oscillator type for phason branch leads to the $\omega^{-1/2}$
dependence of longitudinal relaxation rate \cite{Zumer,Zeyher}.
At the same time,
the theory also derives positive proportionality of $T_1^{-1}$ to
temperature.
Then,
this theory can be ruled out,
because a decrease of $T_1^{-1}$ was observed with an increase of
temperature as shown in Fig.\ref{fig:T1H_log}.

The second theory describes the dynamical susceptibility based on a 
one-dimensional random walk model (details of this model are given
by the reference).
More generally,
the frequency dependence of $T_1^{-1}$ is different for one-, two-,
and three-dimensional diffusive motions \cite{McDowell,Kimmerle},
\begin{align}
&\text{1D:} \quad T_1^{-1} = A \tau_c^{1/2} \omega^{-1/2} \\
&\text{2D:} \quad T_1^{-1} = B \tau_c \ln(\omega) \\
&\text{3D:} \quad T_1^{-1} = C \tau_c - E \tau_c^{3/2} \omega^{1/2}.
\end{align}
The proportionality constants $A$, $B$, $C$, and $E$ depend on the
particular model of hopping motion on a given network of atomic 
sites.
It is clear from Fig.\ref{fig:freq_depend_all} that one-dimensional
diffusive motion exists in PPP at our measuring frequencies.
The following equation is obtained as a conclusion;
\begin{align}
&T_1^{-1} = M_2 f(\omega) \label{1D}\\
&f(\omega) = \frac{1}{\sqrt{2}} \tau_c^{1/2} \omega^{-1/2} \quad \text{(if $\omega < \tau_c^{-1}$)},
\end{align}
where $f(\omega)$ is the correlation function,
$\tau_c$ is the correlation time identical to the inverse of the
diffusion rate,
and $M_2$ is the second moment of the interaction that affects
the relaxation.

In the case of undoped {\it trans}-polyacetylene,
one-dimensional electron hopping affects $^1$H longitudinal
relaxation time \cite{Nechtschein1980,Nechtschein1983}.
It is difficult in the present study to estimate the second moment
by taking into account electron spin-spin interaction and hyperfine
contribution,
because density of free electron is not known.
If mobile electrons govern the $T_1$ measurements,
the relaxation rate should become larger with an increase in temperature\cite{Nechtschein1983}.
As shown in Fig.\ref{fig:T1H_log},
such a dependence was, however, not observed for PPP,
leading us to the assumption that the spin relaxation 
is dominated by not electron spins but fluctuation by molecular motions;
here, one-dimensional diffusive(or propagating) motion of backbone twist is the most plausible candidate.
Moreover,
the $^1$H longitudinal relaxation rate ($T_1^{-1}$) at 303K,
333K, and 363K as a function of the frequency are
shown in Fig.\ref{fig:freq_depend_all}b-d,
respectively.
The relaxation rates in this temperature range show the dependence of
$T_1^{-1} \propto \omega^{-0.5}$.
Therefore we can say that in the wide temperature range including 295 K
there are the one-dimensional diffusion-like twist motion of phenylene
rings along the chain.


\subsection{$^1$H Spin-lattice Relaxation Time and Correlation Time}

\begin{figure}[htbp]
\begin{center}
\includegraphics*[scale=0.4]{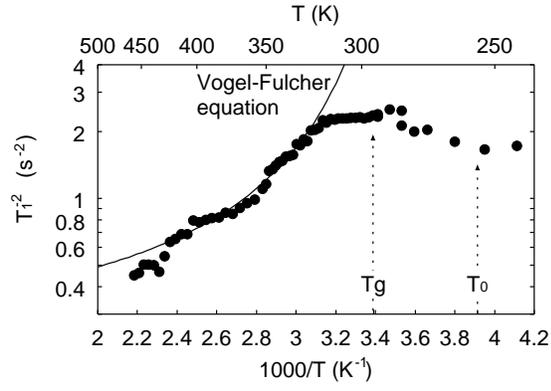}
\caption{The plot of 400 MHz $^1$H longitudinal relaxation rate $(T_1^{-1})^2$ as a function of 1000/T. When one-dimensional diffusion is responsible for the behavior of $T_1^{-1}$, $(T_1^{-1})^2$ is proportional to $\tau_c$.}
\label{fig:T1H_VF}
\end{center}
\end{figure}

If the diffusive motion keeps one-dimensionality in our measuring
temperature range,
\begin{align}
(T_1^{-1})^2 \propto \tau_c
\end{align}
is valid from Eq.\ref{1D}.
As Fig.\ref{fig:T1H_VF} shows,
the correlation time in the temperature range above 295 K follows
the Vogel-Fulcher-Tamman equation\cite{VFT}
\begin{align}
(T_1^{-1})^2 \propto \tau_c = \tau_0 \exp \left( \frac{B}{T-T_0} \right)
\end{align}
where B is constant and the Vogel-Fulcher-Tamman temperature ($T_0$) is
found to be 255.4 K.
The temperature $T_0$ is usually determined by analyzing the
temperature dependence of the correlation time $\tau_c$ of the
so-called $\alpha$ process using the Vogel-Fulcher-Tamman equation.
The Vogel-Fulcher-Tamman temperature $T_0$ is observed at roughly 50 K below
$T_g$ for most amorphous polymers \cite{Ferry}.
Therefore,
taking into account of three factors:
(i) the specific heat jump without latent heat near 295 K,
(ii) the one-dimensional diffusion-like twist motion in the wide temperature
range including 295 K,
and (iii) the correlation time following the Vogel-Fulcher-Tamman equation in the
temperature range above 295 K,
we can explain that within the amorphous PPP the twist motions between
adjacent phenyl rings are frozen at below 295 K without thermodynamic
stability.

There is an interesting prediction to check our explanation.
For hard ellipsoids of revolution M.Letz {\it et al.} have calculated
the phase diagram for the idealized glass transition \cite{Letz}.
They obtained three types of glass transition line.
The first glass transition is the conventional one for
spherical particles driven by the cage effect.
At the second transition,
which occurs for rather nonspherical particles,
a glass phase is formed that consists of domains.
Within each domain there is a nematic order where the center of mass
motion is quasiergodic,
whereas the interdomain orientations build an orientational glass.
The third transition line occurs for nearly spherical ellipsoids
where the orientational degrees of freedom with odd parity, e.g.,
180$^\circ$ flips, freeze independently from the positions.
Again, 
it is widely accepted that PPP is a thermally stable material \cite{Kovacic1987,Yoshino,Ballard}.
In our experiments,
a conventional glass transition dominated by the cage effect was unable to exist
because of high thermal stability of PPP,
whereas a glass transition caused by frozen order parameter relating
to the twist motions between adjacent phenyl rings can occur far below 
the conventional glass transition temperature.

One thing we must stress is the fact that
whereas the temperature dependence of $T_1^{-1}$ shows the typical Vogel-Fulcher-Tamman behavior, the decay profile of longitudinal relaxation measurements
(Fig.\ref{fig:T1_fit}) did not show any evidences to the existence of 
heterogeneity of the glassy structure, namely, distribution in correlation 
time.
In most glassy solids, distribution in the correlation time was realized 
and accounded by modifying the correlation function with 
the extended exponential function(Kohlrausch-Williams-Watts function)\cite{KWW}
 or by using modifying the spectral density function with 
Cole-Cole,\cite{CC}
Cole-Davidson,\cite{CD}
Havriliak-Negami types,\cite{HN}
and so on.
From Fig.\ref{fig:T1H_log}, 
the critical slowing down was observed 
at temperature over the range from $T_g$ to $T_g+20$.
This indicates that the structural relaxation rate($\tau(0)^{-1}\sim \tau_c^{-1}$) is
smaller than the resonance frequency(400MHz);
indeed, more precise estimation of $\tau(0)^{-1}$ is presented 
in Fig.\ref{fig:T1HT2_OPF_critical}.
The existence of the slower structural relaxation rate than the resonance frequency
we used in this paper,
must induce a singularity at the angular frequency of $\omega = \tau(0)^{-1}$  
in the spectral density function,
which should lead to the change in the power law.
We did not observed such a change over the entire temperature range we made
measurements;instead, the $\omega^{-1/2}$ dependence was almost preserved.
We describe the alternative explanation for the preservation of the power law in the following.

Assuming the one-dimensional fluctuation instead of the isotropic one
(Debye model) and distribution of the correlation time,
the spectral density function with a discrete correlation time distribution 
becomes to be the following form:
\begin{eqnarray}
f(\omega)&=& \frac{1}{\sqrt{2}}(p_0\tau_{0c}^{1/2} + p_1\tau_{1c}^{1/2} + \cdots)\omega^{-1/2}\\
&=& \frac{1}{\sqrt{2}}\{\sum_i p_i \tau_{ic}^{1/2}\}\omega^{-1/2},
\end{eqnarray}
where $p_i$ is a probability of which structural relaxation time is $\tau_{ic}$.If the correlation time distribution is continuous,
\begin{eqnarray}
f(\omega) &=& \frac{\omega^{-1/2}}{\sqrt{2}}\int^{\infty}_0 p(\tau) \tau^{1/2} d\tau,\label{SDF}
\end{eqnarray}
where $p(\tau)$ is a continuous version of $p_i$.
From Eq.\ref{SDF}, 
the $\omega^{-1/2}$ law is preserved irrespective of the structural heterogeneity and the $^1$H-$^1$H spin diffusion,
by virtue of separation of variables, $\omega$ and $\tau$.
This situation is realistic when the one-dimensional modulation waves
of backbone twist in PPP has a dispersion.


\subsection{$^1$H Spin-lattice Relaxation Time and Critical Behavior}

Although the validity of the concept of dynamic scaling in glassy solids 
is a controversial problem in modern physics,
it would be meaningful to investigate the critical dynamics of 
glass transition by using knowledges obtained by studies of 
the second order phase transition.

In this section, we analyze the results of the longitudinal relaxation time 
by a theory originally developed for incommensurately modulated solids.\cite{Decker} 
Here,
we restrict ourselves only to make brief introduction of the theory.

By assuming a direct relaxation process of the nuclear spin system for
the case in which the relaxation mechanism is due to the order parameter
fluctuations and by considering the classical fluctuation-dissipation theorem,
one can derive the critical contribution from the following:
\begin{align}
\frac{1}{T_{1cr}}\propto W_1\propto T\sum_k\frac{\chi''({\bf k},\omega)}{\omega},
\label{eq:sum}
\end{align}
where ${\bf k}={\bf q}-{\bf q}_p$ is the reduced wave vector and
${\bf q}_p$ is the wave vector of the periodicity of the polymer chain.
Here assuming that the amorphous portion of the polymer has 
the plate-like, linear conformation because of the relatively small molecular weight
(DP=38-76),
it can be thought that there exists the structural periodicity along the chain.
This assumption seems to be adequate because the one-dimensional fluctuation
is observed by the frequency dependent longitudinal relaxation measurements.
If we think
a model system of interacting pseudospins in which the interaction of
the spins with the lattice vibrations induces backbone twist 
of the phenylene ring of PPP,
the dynamical susceptibility can be written in the following:
\cite{Bonera,Suzuki}
\begin{align}
\chi({\bf k},\omega)=\frac{\chi({\bf k},0)}{1+i\omega\tau({\bf k})},
\end{align}
where $\chi({\bf k},0)$ is the static order parameter susceptibility.
$\chi({\bf k},0)$ depends on the wave vector according to a generalized
Ornstein-Zernike representation \cite{Decker}
\begin{align}
\chi({\bf k},0)=\frac{\chi(0,0)}{1+({\bf k}\xi)^{2-\eta}},
\end{align}
where $\xi$ is the correlation length.
In the critical region, 
it depends on the temperature
with a critical exponent $\nu$ ($\xi\propto|T-T_g|^{-\nu}$).
$\eta$ is a critical exponent defining a small correction from
the Ornstein-Zernike behavior at the glass transition.
Supposing the validity of the van Hove theory for critical dynamics,
we can obtain the following formula for the fast motion regime
($\Omega = \omega \tau(0) \gg 1$):
\begin{align}
\frac{1}{T_{1cr}}\propto T^2e^{E_a/k_BT}\frac{\chi^2(0,0)}{\xi^3}.
\end{align}
Since it may be assumed the critical behavior of the quantities
$\xi$ and $\chi(0,0)$ to diverge as $(T-T_g)^{-\nu}$ and 
as $(T-T_g)^{-\gamma}$ near $T_g$, respectively,
the temperature dependence of the critical behavior is reflected by
the critical exponent $\zeta = 2\gamma-3\nu$.
Therefore, the final formula can be obtained in the following:
\begin{align}
\frac{1}{T_{1cr}}\propto T^2e^{E_a/k_BT}(T-T_g)^{-\zeta}.
\label{eq:zeta}
\end{align}
where $E_a$ is an activation energy for Arrhenius-type thermally activated 
process without interaction and the other symbols are the usual meaning.
For the slow-motion regime ($\Omega \ll 1$),
the $1/T_{1cr}$ will become temperature independent.

The temperature dependence of 400MHz $^1$H $T_1$ is shown in Fig.\ref{fig:T1H_log} over a broad temperature interval including $T_g$.
Since the $^1$H longitudinal relaxation is mainly dominated by nuclear magnetic dipole
interaction,
we supposed  non critical background relaxation mechanism 
by a two-phonon Raman process as a negligible effect.\cite{Abragam}
Therefore,
the critical relaxation rate $1/T_{1cr}$ can be evaluate from the
$T_1$ data ($1/T_1\approx1/T_{1cr}$) for a wide range of
temperature above $T_g$.

The prefactor $T^2\exp(E_a/k_BT)$ in Equation (\ref{eq:zeta}) depends
noncritically on the temperature.
In the case of PPP,
$E_a$ corresponds to the activation energy to promote the molecule
from a nonplanar to a planar state.
S.Guha {\it et al.} obtain an activation energy $E_a = 0.04$ eV
for {\it para} hexaphenyl via Raman scattering \cite{Guha}.
They also calculated $E_a$ from the difference in internal energy
between the planar and nonplanar configurations of a biphenyl
molecule \cite{Guha}.
The Hartree-Fock method yields $E_a = 0.145$ eV compared to the
density functional method which yields $E_a = 0.089$ eV.
From the energy calculation of a PPP chain in the reference
\cite{Ambrosch} it is observed that $E_a = 0.065$ eV.
The calculated values for the activation energy are higher for two
reasons:
biphenyl is shorter than hexaphenyl and calculations for both
biphenyl and PPP chain are done for isolated molecules,
where we expect the planarizing forces to be weaker than for
hexaphenyl which is an ensemble of long molecules in a surrounding
environment.
Therefore in the sample of PPP (DP $=$ 38 $\sim$ 76) we uesd 
the activation energy $E_a$ is likely to become smaller than 0.04 eV.
Thus,
the critical temperature dependence of $T_{1cr}$ can be represented
by a critical exponent $\zeta$ according to
\begin{align}
\frac{1}{T_{1cr}}\approx \frac{1}{T_1}\propto T^2(T-T_g)^{-\zeta}.
\end{align}
Here,
we omit the prefactor $\exp(E_a/k_BT)$ which has a negligible effect
on the fit value $\zeta$ in a temperature interval above $T_g$.

\begin{figure}[htbp]
\begin{center}
\includegraphics*[scale=0.6]{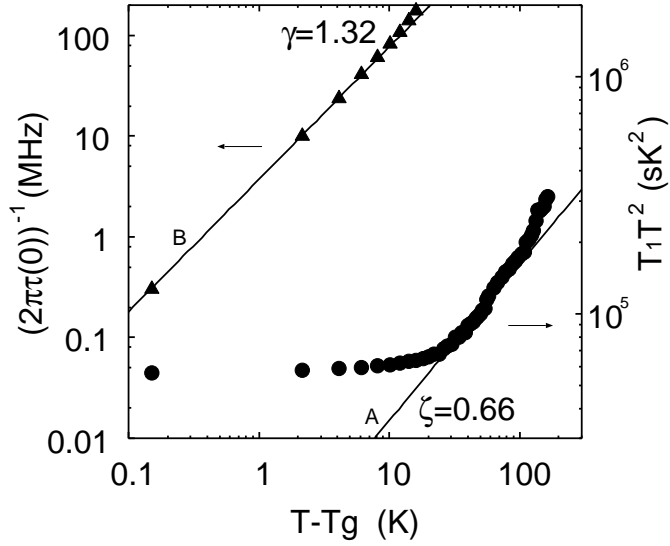}
\caption{The plot of $T_1T^2$ as a function of $T-T_g$ (right scale). Moreover, the plot of the order-parameter relaxation frequency $(2\pi\tau(0))^{-1}$ as a function of $T-T_g$ (left scale).}
\label{fig:T1HT2_OPF_critical}
\end{center}
\end{figure}

Figure \ref{fig:T1HT2_OPF_critical} shows
the critical temperature dependence of $T_{1cr}$ ($\approx T_1$).
At temperatures from $T_g$ to $T_g$+10 
the slow-motion limit behavior is observed,
where $T_1T^2$ is temperature independent.
Over $T_g$+20 one reaches the fast-motion limit, 
where $T_1T^2$ become to show a power law with a critical
exponent $\zeta = 0.66$.
For $T > T_g+60$ the value of $T_1T^2$ show a steeper increase
with an increase of temperature.
This effect is similar to that observed by Decker {\it et al.}\cite{Decker}
and may be due to the decrease of the correlation length with an increase of 
temperature.

This value for the experimental critical exponent $\zeta = 0.66$ 
can be compared with the theoretical exponent $\zeta = 0.663$ 
for the three-dimensional XY model derived from
an elaborated field-theoretical model that goes beyond the mean-field
approach \cite{Kaufmann}.
 
The critical exponents $\zeta = 2\gamma-3\nu$ derived for the 3D Ising,
XY,
and Heisenberg models are 0.5904 $\pm$ 0.0042,
0.6241 $\pm$ 0.0054,
and 0.6570 $\pm$ 0.0057,
respectively \cite{Gebhardt}.
However,
the critical exponent $\zeta = 0.66$ derived above from our experiments
differs from the prediction of the 3D XY model.
Thus,
the applicability of both other models cannot be strictly ruled out.

The relaxation time $\tau(0)$ of the order parameter can be 
derived by following equation \cite{Decker}
\begin{align}
\left(\frac{T_1}{T_1^{fml}}\right)^2=\frac{1}{2}(\sqrt{\Omega^2+1}+1).\label{taum1}
\end{align}
For $T_1T^2$ we employed the experimental data in Fig.\ref{fig:T1HT2_OPF_critical}
while $T_1^{fml}T^2$ is calculated by using the line A.
In Fig.\ref{fig:T1HT2_OPF_critical} also the order parameter
relaxation frequency $(2\pi\tau(0))^{-1}$ calculated by using Eq.\ref{taum1}
is plotted versus $(T-T_g)$ on a double logarithmic scale.
The straight line B is a fit of the data to a power law yielding
a critical exponent of $z\nu = 1.32$,
which fits very well to the static susceptibility exponent
$\gamma=1.3160\pm0.0012$ predicted for the 3D XY model
\cite{Gebhardt}.
Comparison with the corresponding theoretical $\gamma$ values for the
3D Ising and Heisenberg models,
1.2402 and 1.3866,
respectively.
From the above analysis,
the 3D-XY model is a plausible universaility class for
the critical dynamics of twist-glass transition in PPP.


\section{Conclusion}
Dynamics of Yamamoto-type poly(p-phenylene)[PPP] was investigated by differential scanning calorimetry(DSC) and proton solid-state NMR relaxation spectroscopy.
The DSC chart shows the baseline jump without latent heat at 295K,
which is due to the twist glass transition of the polymer. 
From the variable temperature proton longitudinal relaxation time($T_1$)
measurements,
we observed relatively short $T_1$ over the wide range from 250K
(closed to Vogel-Fulcher-Tamman temperature) to 360K, 
inferred the existence of cooperative critical slowing down associated with 
the glass transition.
The frequency dependence of proton longitudinal relaxation time at 295-363K
shows $R_1 \sim \omega^{-0.5}$ dependence,
which is due to the one-dimensional diffusion-like motion of the backbone
conformational modulation.
The dependence indicates that closed to $T_g$ from the higher temperature 
the cooperative rearrangement region(CRR) grows in the one-dimensional manner.
From the pseudospin argument, 
the 3D-XY model is a plausible universality class
for the critical dynamics of twist-glass transition in PPP.
This research is an example that have obtained the information about
the universality class of twist-glass transition found in $\pi$-conjugated 
polymers.


\section*{Acknowledgments}
This work is supported by Ministry of Education, Culture, Sports, Science
and Technology(Japan) through a Grant-in-Aid for science research
(No.16685012).
The authors are grateful to Professor Tetsuo Asakura and Dr.Kazuo Yamauchi
(Tokyo University of Agriculture and Technology) for fruitful discussions. 



\newpage

\end{document}